# Cosmic history and a Candidate Parent Asteroid for the Quasicrystal-bearing Meteorite Khatyrka

Matthias M. M. Meier[1]*, Luca Bindi[2,3], Philipp R. Heck[4], April I. Neander[5], Nicole H. Spring[6,7], My E. I. Riebe[1,8], Colin Maden[1], Heinrich Baur[1], Paul J. Steinhardt[9], Rainer Wieler[1] and Henner Busemann[1]

[1]Institute of Geochemistry and Petrology, ETH Zurich, Zurich, Switzerland. [2]Dipartimento di Scienze della Terra, Università di Firenze, Florence, Italy. [3]CNR-Istituto di Geoscienze e Georisorse, Sezione di Firenze, Florence, Italy. [4]Robert A. Pritzker Center for Meteoritics and Polar Studies, Field Museum of Natural History, Chicago, USA. [5]Department of Organismal Biology and Anatomy, University of Chicago, Chicago, USA. [6]School of Earth and Environmental Sciences, University of Manchester, Manchester, UK. [7]Department of Earth and Atmospheric Sciences, University of Alberta, Edmonton, Canada. [8]Current address: Department of Terrestrial Magnetism, Carnegie Institution of Washington, Washington, USA. [9]Department of Physics, and Princeton Center for Theoretical Science, Princeton University, Princeton, USA.

*corresponding author: matthias.meier@erdw.ethz.ch; (office phone: +41 44 632 64 53)

## Abstract

*The unique CV-type meteorite Khatyrka is the only natural sample in which "quasicrystals" and associated crystalline Cu,Al-alloys, including khatyrkite and cupalite, have been found. They are suspected to have formed in the early Solar System. To better understand the origin of these exotic phases, and the relationship of Khatyrka to other CV chondrites, we have measured He and Ne in six individual, ~40-µm-sized olivine grains from Khatyrka. We find a cosmic-ray exposure age of about 2-4 Ma (if the meteoroid was <3 m in diameter, more if it was larger). The U,Th-He ages of the olivine grains suggest that Khatyrka experienced a relatively recent (<600 Ma) shock event, which created pressure and temperature conditions sufficient to form both the quasicrystals and the high-pressure phases found in the meteorite. We propose that the parent body of Khatyrka is the large K-type asteroid 89 Julia, based on its peculiar, but matching reflectance spectrum, evidence for an impact/shock event within the last few 100 Ma (which formed the Julia family), and its location close to strong orbital resonances, so that the Khatyrka meteoroid could plausibly have reached Earth within its rather short cosmic-ray exposure age.*

## 1. Introduction

The motivation for this noble gas study is the curiosity and fascination with the origin of an exotic type of material: quasicrystals. Short for quasi-periodic crystals, they are materials showing a quasi-periodic arrangement of atoms, and rotational symmetries forbidden to ordinary crystals (e.g., five-fold). First proposed by Levine and Steinhardt (1984), they were first synthesized and identified in the laboratory by Shechtman et al. (1984). The ensuing multi-decade search for natural quasicrystals was eventually successful when a powder from a millimeter-sized rock sample (later called the "Florence Sample") from the collection of the Museum of Natural History in Florence, Italy, displayed a diffraction pattern with a five-fold symmetry (Bindi et al., 2009). This first natural quasicrystal, with a composition $Al_{63}Cu_{24}Fe_{13}$, was named icosahedrite. In the host rock of the Florence sample, silicates and oxides are partially inter-grown with exotic copper-aluminium-alloys (khatyrkite, $CuAl_2$, and cupalite, CuAl) which contain the quasicrystals. The oxygen isotopic compositions of silicates and oxides in the Florence Sample suggest an extraterrestrial origin (Bindi et al., 2012), as they match the ones found in some carbonaceous chondrites. The provenance of the Florence Sample was eventually traced back to a Soviet prospecting expedition to the Far East of Russia in 1979 (Bindi and Steinhardt, 2014). To find more of the exotic material, an expedition was launched to the Koryak mountain range in the Chukotka Autonomous Region in 2011. Eight millimeter-sized fragments of extraterrestrial origin were found by panning Holocene (>7 ka old) river deposits (MacPherson et al., 2013). Again, Cu,Al-alloys were found attached to, or inter-grown with, extraterrestrial silicates and oxides. The oxygen isotopes, chemistry and petrology of the eight new fragments all suggest a $CV_{ox}$ (carbonaceous chondrite, Vigarano-type, oxidized subtype) classification for the meteorite, now named Khatyrka (Ruzicka et al., 2014). The achondritic, diopside-hedenbergite-rich and Cu,Al-alloy-encrusted Florence Sample suggests that Khatyrka is an unusual chondritic breccia which accreted both achondritic and exotic materials (MacPherson et al., 2013).

So far, no other meteorite is known to contain Cu,Al-alloys or quasicrystals. The lack of $^{26}Mg/^{24}Mg$ anomalies suggests that the Cu,Al-alloys formed after the primordial $^{26}Al$ ($t_{1/2}$ = 0.7 million years, Ma) had decayed away, i.e., at least ~3 Ma after formation of the oldest condensates in the Solar System 4.567 billion years (Ga) ago (MacPherson et al., 2013). Hollister et al. (2014) found high-pressure mineral phases (including stishovite and ahrensite) indicating Khatyrka was at one point exposed to pressures >5 Gigapascals (GPa) and temperatures >1200 °C, followed by rapid cooling. Synthetic icosahedrite remains stable under these conditions (Stagno et al., 2015). These conditions are also typical for asteroid collisions (Stoeffler et al., 1991). In dynamic shock experiments, Asimow et al. (2016) successfully synthesized icosahedral phases with a composition close to that of icosahedrite, and Hamann et al. (2016) reported the shock-induced formation of khatyrkite ($CuAl_2$), mixing of target and projectile material, and localized melting along grain boundaries, resembling the assemblages found in Khatyrka. Lin et al. (2017) argued that a two-stage formation model is needed to explain the mineralogical and petrographic features observed in Khatyrka: first, an event as early as 4.564 Ga in which quasicrystals with icosahedrite composition formed; and, second, a more recent impact-induced shock that led to the formation of a second generation of quasicrystals, with a different composition. Ivanova et al. (2017) suggested the possibility of an anthropogenic origin of the Cu,Al-alloys from mining operations, but this explanation is incompatible with the chemistry and the thermodynamic conditions needed to explain all observations listed above (see also MacPherson et al., 2016).

In this work, we aim to better understand the origin and history of the quasicrystal-bearing Khatyrka meteorite and its relationship to other meteorites, in particular other CV chondrites. We do this by measuring the helium and neon (He, Ne) content of individual Khatyrka olivine grains, in order to determine their cosmic-ray exposure and radiogenic gas retention (based on uranium-thorium-helium = U,Th-He) ages. Due to the extremely small mass available from Khatyrka (total mass <0.1 g), destructive analy-ses have to be reduced to the minimum. We worked with single grains of ca. 40 μm diameter, which were all part of a chondrule from Khatyrka fragment #126 (recovered in the 2011 expedition). Because of the very small gas amounts expected, we used the high-sensitivity "compressor-source" noble gas mass spectrometer at ETH Zurich (Baur, 1999). This unique instrument has previously been used in a similar way to measure He and Ne in individual mineral grains returned from asteroid Itokawa (Meier et al., 2014).

## 2. METHODS

### 2.1. Volumes from X-ray tomography

To determine cosmic-ray exposure and radiogenic gas retention ages, we need noble gas concentrations ($cm^3$STP/g), which requires the determination of the masses of the individual grains. This is not possible to do both safely and reliably on a micro-balance for such small (<1 μg) grains. To determine their masses, we first measure their volumes using high-resolution X-ray tomography (nano-CT), and then multiply these volumes with the densities calculated from their mineralogical composition (determined with SEM-EDS) and textbook mineral densities. At the University of Florence, seven olivine grains from Khatyrka fragment #126 (named #126-01 through -07, or 1 through 7 for short in the figures), all fragments from a chondrule (judging from their size and chemistry), were transferred with a micro-manipulator needle from a TEM grid to a carbon tape. The samples were then imaged (on the tape) with a GE manufactured v|tome|x s micro/nano-CT scanner located at the UChicago PaleoCT facility, using the 180kV nano-CT tube, an acceleration voltage of 80 kV, beam current of 70 μA with 500 ms integration time, and no filter. The voxels of the scan are isometric with a size of 5.473 μm. Grain volumes were then determined from the CT images using the "3D-object counter" plug-in in Fiji/ImageJ (Schindelin et al., 2012), which finds 3D-connected objects with a voxel brightness above a user-defined 8-bit (256) gray-scale value (called the "threshold"). We first searched a range of thresholds where the grains are well-resolved from the background, then searched for the sequence of four thresholds for which the volume change between adjacent gray-scale steps was minimal. We then corrected for surface resolution effects (i.e., for voxels close to the actual grain which are only partially in-filled by grain material, and thus have a reduced voxel brightness: they fall below the threshold although they contain a sub-voxel-sized part of the grain) by interpolating between the average brightness of unambiguous (internal) grain voxels and background voxels (e.g., if the grain material brightness is 120 and the background brightness is 40, a surface-near voxel with brightness 80 is assumed to be half in-filled with grain material). Grain #126-06 was too small, and too fragmented to be of use for this study, and was thus not further analyzed. The grain volume uncertainties in Table 1 correspond to the range in volume within the sequence of four thresholds with minimal volume changes (after correction for surface effects). Grain #126-05 fragmented after nano-CT scanning (as visible from the SEM images, see section 2.2.). Only two of its fragments could be transferred to the sample holder for noble gas analysis. Their volume was estimated from their cross-sectional area in the SEM images and an empirical relationship between cross-sectional area (A) and volume (V) established by the other five grains, A = (0.080±0.009) × V ($R^2$ = 0.998), which resulted in a somewhat larger mass uncertainty for these fragments. The volume of grain #126-05 given in Table 1 represents the sum of the two fragments.

### 2.2. Chemical composition, grain masses

After nano-CT scanning, the bulk elemental composition of the grains was qualitatively de-termined by spot analysis on unpolished surfaces using an Oxford Instruments XMax-50 energy dispersive X-ray (EDX) spectrometer, mounted on a Zeiss Evo 60 Scanning Electron Microscope (SEM; acceleration voltage 20 kV, electron current 400 pA) at the Field Museum of Natural History in Chicago. Back-Scattered Electron (BSE) images of the grains are shown in Fig. S1 in the Supplementary Material. We found an average composition of 95±5% forsterite ($Mg_2SiO_4$; 3.28 g/cm$^3$) and 5±5% fayalite ($Fe_2SiO_4$; 4.39 g/cm$^3$), corresponding to a grain density of 3.33±0.06 g/cm$^3$ (Table 1). Since the individual measurements were done in spot analysis mode and are not resolved from their average, we only use the average density to calculate the grain masses in Table 1. The six grains (or grain fragments) were then transferred to a customized Al sample holder for noble gas analysis, using a hydraulic Narishige MMO-202ND/MMN-1 micro-manipulator attached to a Nikon Eclipse Microscope at SEAES, University of Manchester (an image of a grain in transfer is shown in Fig. S2 in the Supplementary Material).

## 2.3. Light noble gas analysis

At the noble gas laboratory of ETH Zurich, the sample holder with the grains was loaded into a sample chamber which remains connected to an ultra-low blank extraction line and the "compressor-source" noble gas mass spectrometer (Baur, 1999) at all times (i.e., no valves are operated during a measurement run). Noble gases were extracted by laser heating using a Nd:YAG laser (λ=1064 nm), focused on each grain for 60 seconds, melting and, in some cases, completely vaporizing the grains. Each extraction was monitored using a camera, which showed the intense flickering / glowing of the grains during laser heating. In two cases (grains #126-02 and -05), some residual light emission was present after 60 seconds, suggesting incomplete degassing. These grains were thus extracted again for another 60 seconds in a follow-up analysis run, and the cumulative gases released by each grain were determined in post-processing. After extraction, the released gases were passed through the extraction line which features two getters and three cold traps (two of them containing activated charcoal) cooled with liquid nitrogen. The strong increase in sensitivity provided by the compressor source also leads to a very fast removal of Ar, Kr, and Xe due to ion pumping, and hence only He and Ne can be analyzed. Isotopic ratios and elemental abundances of $^3$He, $^4$He, $^{20}$Ne,

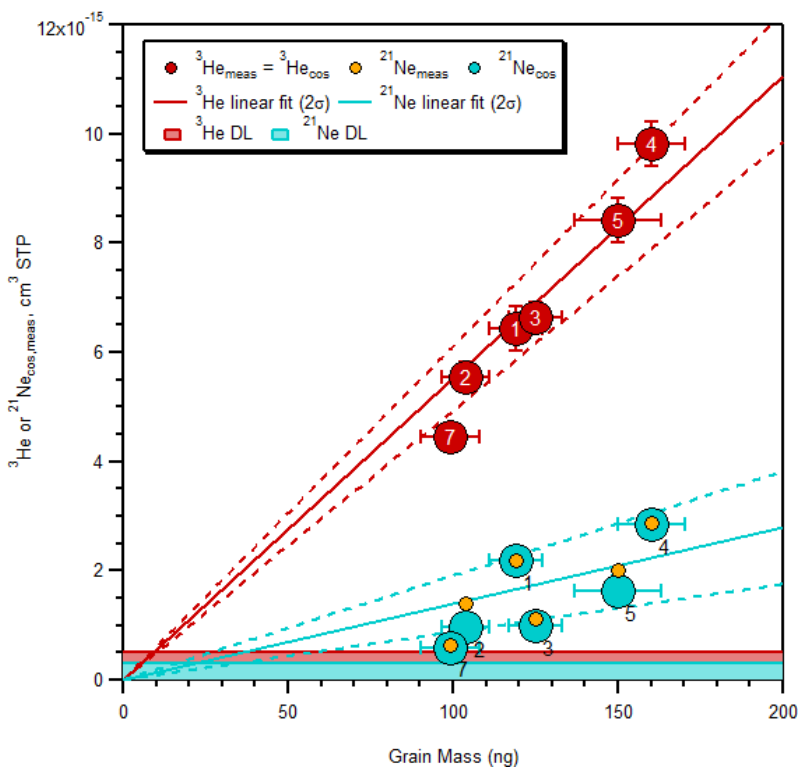

***Fig. 1: cosmogenic $^3$He and $^{21}$Ne in Khatyrka olivine grains.*** *The measured, purely cosmogenic $^3$He (red symbols), measured $^{21}$Ne (orange symbols) and cosmogenic $^{21}$Ne (blue symbols) gas amounts (units of 10-15 cm3STP) are plotted against grain mass (all errors are 1σ). Both linear fits (lines of constant concentration) are forced through the origin, and are shown with their 2σ confidence intervals (dashed lines).*

$^{21}$Ne, and $^{22}$Ne were then measured (together with HD, H$_2$$^{16}$O, $^{40}$Ar, and CO$_2$ for interference corrections, which proved to be negligible) following a protocol developed by Heck et al. (2007). In this work, we used a new, slightly modified data fitting routine which is only applicable to very low gas amounts, but results in smaller uncertainties in the measured gas amounts and, in some cases, also to somewhat lower detection limits (a detailed discussion is given in the Supplementary Material). A series of blank runs – indentical to the sample runs except that no laser is fired – was measured to determine the extent their scattering around their average. Two standard deviations of that scatter added to the average blank define the detection limits for all isotopes, which are (in units of 10$^{-15}$ cm$^3$STP; 1 cm$^3$STP = 2.687×10$^{19}$ atoms) 0.55, 310, 38, 0.47, and 3.9 for $^3$He, $^4$He, $^{20}$Ne, $^{21}$Ne, and $^{22}$Ne, respectively. The average blanks represent 8%, 19%, 21%, 34%, and 36% of the detection limits for these isotopes, respectively. A calibration bottle containing both air-like He, and near-atmospheric Ne (Heber et al., 2009) was used to determine instrumental sensitivity for $^3$He, $^4$He, and $^{20}$Ne, and the instrumental mass fractionation for Ne isotopes (0.17±0.06‰/amu, based on the measured $^{20}$Ne/$^{22}$Ne ratio). A significant excess signal (ca. 8%) on mass 21 was found in calibration runs, which was identified as interference from $^{20}$NeH. The $^{20}$NeH/$^{20}$Ne ratio needed to explain the interference in the calibrations runs is 0.24±0.01‰. The H pressure in the spectrometer, monitored via measurement of HD, stayed constant over the course of the measurement run, implying that the NeH interference remained roughly constant. The interference was subsequently subtracted from mass 21 in the sample runs. That correction was never more than 3% of the total signal on mass 21.

*2.4. Orbital evolution model*

As the orbital evolution model used here to determine the possible origins of the Khatyrka meteoroid in the asteroid belt was presented in Meier et al. (2017), we will only give the model parameters here. To determine the migration range of the Khatyrka meteoroid, we evolved the semi-major axes of test particles placed 0.004 Astronomical Units (au) in- and outside the $v_6$, 3:1, and 5:2 resonances backwards, using the formulas given by Vokrouhlicky et al. (2015) for seasonal and diurnal Yarkovsky drift, and Farinella et al. (1998) for occasional resets of rotation rates and obliquity by collisions. Test particles were given five different sizes: 0.5, 1.0, 1.5, 3.0, and 5.0 m, a compressed density of 3.3 g/cm$^3$ and a porosity of 22% (corresponding to bulk CV chondrites; the resulting bulk density used in the model is 2.6 g/cm$^3$), an albedo of 0.15, a heat capacity of 500 J kg$^{-1}$ K$^{-1}$, and a thermal conductivity of either 0.15 or 1.5 W m$^{-1}$ K$^{-1}$. Orbital evolution was continued until the modeled meteoroid had accumulated the measured cosmic-ray-produced inventory of noble gases (the time for this depends on its radius, see section 3.2). Collisional disruptions were neglected since the collisional life-time for all Khatyrka meteoroids considered here is about five times longer than their cosmic-ray exposure age (Farinella et al., 1998). Gravitational interactions with planets (apart from resonances) and other asteroids are neglected. The resonances are treated in a statistical way, with test particles within 0.004 au of resonances removed with a probability scaled to reproduce the resonance-specific life-times (Gladman, 1997). All test particles ejected by resonances were removed from the simulation, and replacement test particles were started until 10$^4$ individual migration histories had been accumulated for each resonance and meteoroid size.

**3. RESULTS**

*3.1. Noble gas components*

During laser melting, all six Khatyrka olivine grains released $^3$He and $^{21}$Ne well above the detection limit (DL; see Table 1, Fig. 1). Also above DL were: $^4$He in three grains (#126-02, -04 and -05), $^{20}$Ne in three grains (#126-02, -03, and -05), and $^{22}$Ne in four grains (#126-02, -03, -04, -05). For the three grains for which $^4$He was below DL, a lower limit for the $^3$He/$^4$He ratio was calculated from the ratio of the measured $^3$He and the $^4$He DL. These $^3$He/$^4$He ratios (and lower limits) are all a factor of at least ~10 higher than the $^3$He/$^4$He ratio in the solar wind or other trapped components, and at least a factor of ~1000 higher than the $^3$He/$^4$He ratio in the Earth's atmosphere. This suggests that the $^3$He in Khatyrka olivine is dominated (>90%) by the cosmic-ray-induced spallation (or "cosmogenic") component ($^3$He/$^4$He ~0.2; e.g. Leya and Masarik, 2009). Fig. 1 shows that for $^3$He, all six grains plot within 2σ of their mass uncertainty (x-coordinate) on a linear fit forced through the origin. The slope of the line corresponds to an average cosmogenic $^3$He ($^3$He$_{cos}$) concentration of 5.5±0.2×10$^{-17}$ cm$^3$STP/ng (or 10$^{-8}$ cm$^3$STP/g; $R^2$ = 0.993). For the other important cosmogenic isotope, $^{21}$Ne ($^{21}$Ne$_{cos}$), the scatter around the fit line (1.3±0.2×10$^{-17}$ cm$^3$STP/ng; $R^2$ = 0.881) is more pronounced. To calculate the cosmogenic fraction of $^{21}$Ne ($^{21}$Ne$_{cos}$), we use a two-component deconvolution with a non-cosmogenic ($^{21}$Ne/$^{22}$Ne = 0.03; e.g. Ott, 2014) and a cosmogenic end-member ($^{21}$Ne/$^{22}$Ne = 0.89; e.g. Leya and Masarik, 2009). Cosmogenic $^{21}$Ne contrib-utes between 67.5 and 99.9% of the total measured $^{21}$Ne, i.e., the cosmogenic component is dominant also for $^{21}$Ne. The nature of the non-cosmogenic Ne component (e.g., Earth's atmosphere, solar wind, Q gases) in the Khatyrka olivines cannot be determined, due to the high uncertainties in the measured $^{20}$Ne/$^{22}$Ne ratios resulting from the low gas amounts (Fig. S3 in the Supplementary Material). The measured elemental $^{20}$Ne/$^4$He ratios are suggestive of an atmospheric origin (Fig. 2, inset). If only the cosmogenic $^{20}$Ne (=0.92 × $^{21}$Ne$_{cos}$) is considered, all samples plot between cosmogenic and radiogenic end-members in a $^{20}$Ne/$^4$He vs. $^3$He/$^4$He diagram (Fig. 2, main). This suggests that since (and during) Khatyrka's exposure to cosmic-rays, no significant fraction of He was lost, e.g. to impact shocks or solar heating. In summary, He and Ne in Khatyrka olivine are dominated by the cosmogenic component, with a small (<33% for $^{21}$Ne, <10% for $^3$He) contribution of atmospheric (or trapped) gases, and some radiogenic $^4$He.

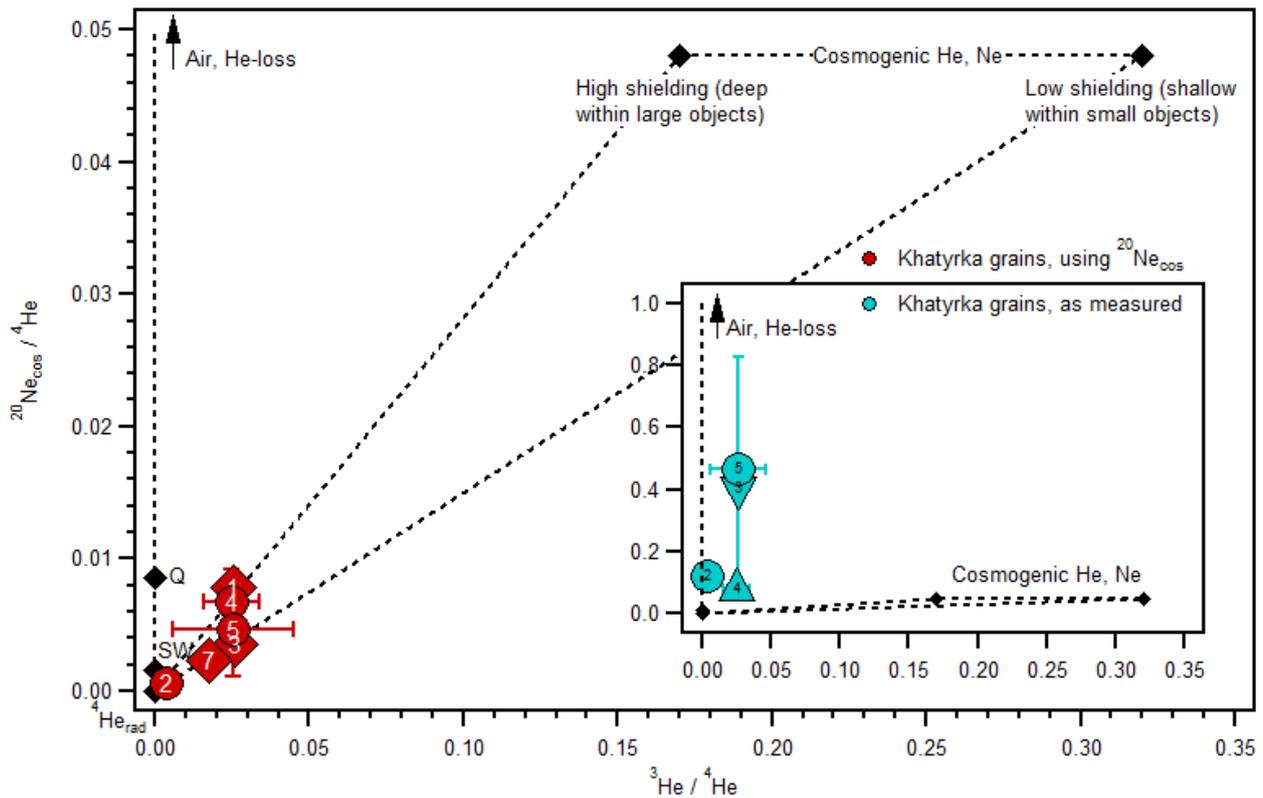

***Fig. 2: $^{20}$Ne/$^4$He vs. $^3$He/$^4$He diagram.*** *Main: The $^{20}$Ne$_{cos}$/$^4$He and $^3$He/$^4$He ratios of the six Khatyrka olivine grains ($^{20}$Ne$_{cos}$ calculated from $^{21}$Ne$_{cos}$). For grains #126-01, -03, and 07, $^4$He was below DL. Therefore, the position of the data point on the diagram corresponds to the minimum distance to the radiogenic end-member (i.e., the origin): these data points are represented by diamonds instead of circles. Inset: Same axes, but expanded to include the measured ratios in $^{20}$Ne/$^4$He (for grains #126-02 and -05), or upper and lower limits on that ratio where $^4$He, or $^{20}$Ne, was below DL (grains #126-03 and -04, respectively; the triangle symbols point towards the unconstrained direction). Noble gas components in both diagrams are shown as black diamonds connected by dashed lines: radiogenic $^4$He at the origin, Q gases (Busemann et al., 2000), and solar wind (SW; Heber et al., 2009), as well as the cosmogenic end-members for low and high shielding (Leya and Masarik, 2009). The terrestrial atmosphere plots outside the diagram at $^{20}$Ne/$^4$He = 3.14, in the same direction as elemental fractionation (loss of He vs. Ne).*

## 3.2. Cosmic-ray exposure age and meteoroid radius

The determination of a cosmic-ray exposure (CRE) age, i.e., the time a meteorite has been exposed to galactic cosmic rays (GCR) as a meteoroid (a meter-sized object in interplanetary space), requires two key values: the concentration of a cosmogenic isotope (e.g., $^3$He$_{cos}$ or $^{21}$Ne$_{cos}$), and a productionrate (atoms per units of mass and time) of that isotope from interactions with target elements. While the former can be determined by measurement, the latter depends on the chemical composition of the meteoroid, its size, and the position of the sample within the meteoroid. The model calculations of Leya and Masarik (2009) allow us to determine He and Ne production rates in forsteritic olivine in a carbonaceous chondritic matrix. This is important because the production of Ne in forsteritic olivine is 60-80% higher compared to a bulk CV chondrite. A "shielding parameter" is needed to reflect the dependence of the cosmogenic production rates on the position within the meteoroid and the size of the meteoroid (together referred to as "shielding"). The most commonly used shielding parameter, the $^{22}$Ne/$^{21}$Ne ratio of the cosmogenic end-member, cannot be used here since it cannot be reliably determined given the large uncertainties in $^{20}$Ne/$^{22}$Ne and the low number of samples (Fig. S3 in Supplementary Materials). Instead, we use the cosmogenic $^3$He/$^{21}$Ne ratio, which – for forsterite – yields CRE ages showing a clear dependence on meteoroid size (Fig. 3). The nominal $^3$He/$^{21}$Ne ratios of the grains vary between ~3 and ~7, encompassing the full range expected from variable shielding. However, since all six grains are derived from the same, mm-sized grain (#126), this variation cannot reflect variable shielding conditions. It does not reflect partial loss of He on the meteoroid or parent body either, as this would result in variable $^3$He and (nearly) constant $^{21}$Ne concentrations, which is the opposite of what is observed (Fig. 1). Instead, the most likely explanation is that the Ne variability reflects incomplete degassing of Ne during laser heating for some of the grains. Indeed, the two grains with the highest $^{21}$Ne$_{cos}$ concentrations (#126-01 and -04) have compatible $^3$He/$^{21}$Ne ratios (of 2.95±0.35 and 3.45±0.33) and $^{21}$Ne$_{cos}$ concentrations (1.83±0.22 and 1.78±0.19 × 10$^{-8}$ cm$^3$STP/g), suggesting that only these two grains were completely degassed for $^{21}$Ne. The error-weighted average of the $^3$He/$^{21}$Ne ratio of all grains is 3.50±0.46 (2σ), and the ratio of the sums of all $^3$He and $^{21}$Ne released is 4.03±0.46 (2σ). The true $^3$He/$^{21}$Ne ratio of Khatyrka is thus most likely in the range of ~3-4 (shaded region in Fig. 3). This range requires that the Khatyrka meteoroid had a radius of at least 30 cm (if the samples are derived from the meteoroid's center), and might have been up to 5 m in radius, or even more, as the latter is just the largest meteoroid modeled by

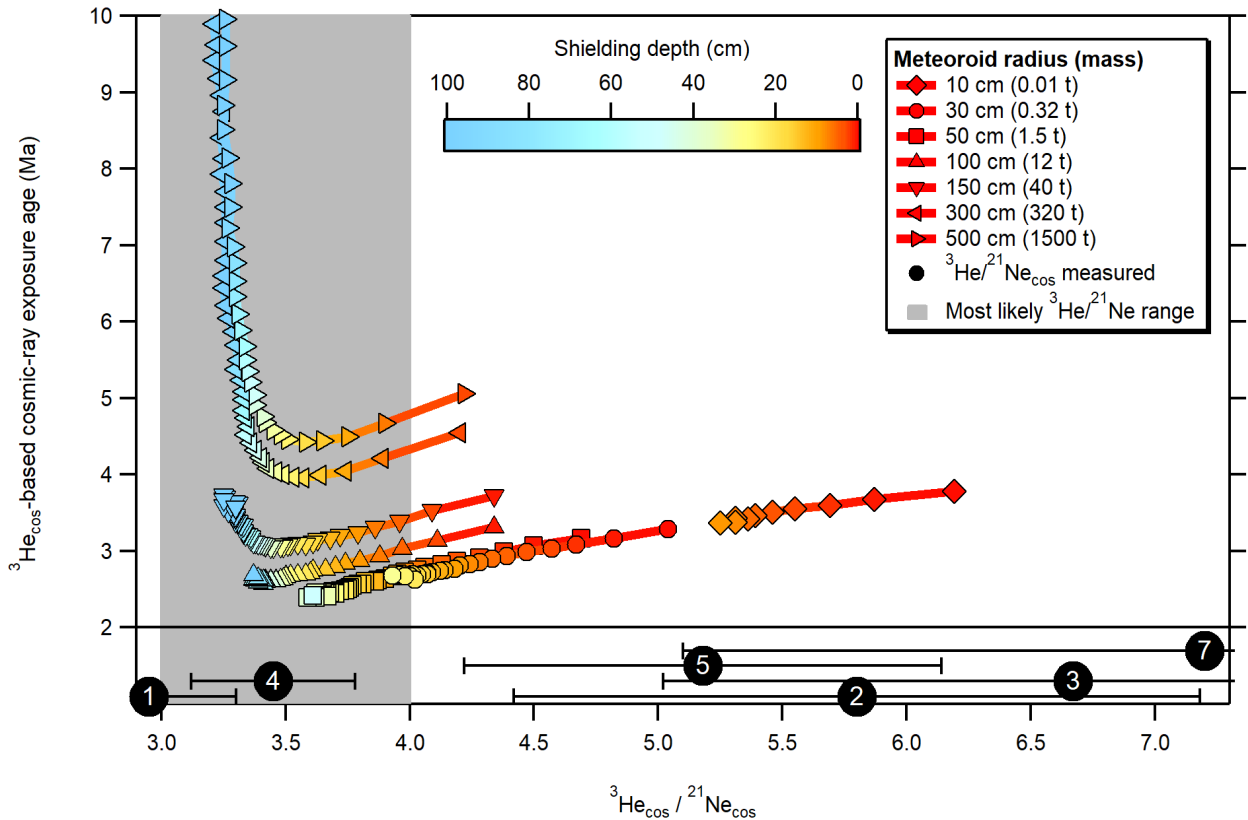

***Fig. 3: CRE age as a function of cosmogenic $^3He/^{21}Ne$, shielding and meteoroid radius.*** *Upper part: the $^3$He-based CRE age of Khatyrka (Leya and Masarik, 2009) as a function of the cosmogenic $^3He/^{21}Ne$ ratio, shielding depth (symbol color) and meteoroid size (symbol shape). Lower part: the $^3He/^{21}Ne$ ratios measured in the individual grains (filled, numbered circles with 1σ errors). The gray shaded area is the range for the error-weighted average of the $^3He/^{21}Ne$ ratio (1σ) for all grains. Reading example: if the true $^3He/^{21}Ne$ ratio of Khatyrka olivine is 3.5, and the meteoroid radius was 150 cm, then the CRE age is ~3 Ma and the shielding depth ~30 cm (light yellow).*

Leya and Masarik (2009). As shown in Fig. 3, the cosmogenic $^3$He concentration (5.5±0.2 × $10^{-8}$ cm$^3$STP/g) and the $^3$He production rates corresponding to a cosmogenic $^3He/^{21}Ne$ ratio in the range ~3-4, result in a likely CRE age range of 2-4 Ma for meteoroids up to about 1.5 m in radius. If the meteoroid was larger than this, and the true cosmogenic $^3He/^{21}Ne$ ratio of Khatyrka olivine is <3.5, the samples could also derive from a strongly shielded position deep (>1 m) inside a large meteoroid, and only a lower limit of >4 Ma can then be given for the CRE age. However, such large meteoroids entering the Earth's atmosphere often result in air-bursts, and thus contribute only a disproportionately small fraction of the flux of meteorites arriving at the surface (Bland and Artemieva, 2006). Therefore, we favor a smaller meteoroid size, and thus a CRE age in the range of 2-4 Ma. A measurement of the cosmogenic radionuclide activities in Khatyrka (e.g., $^{26}$Al and $^{10}$Be), while instrumentally challenging, would significantly improve both the CRE age and pre-atmospheric size determination.

### 3.3. Radiogenic gas retention ages

After subtraction of cosmogenic $^4$He ($^4$He$_{cos}$ / $^3$He$_{cos}$ = 4.5±1.3; Leya and Masarik, 2009), the remaining $^4$He must be from Earth's atmosphere (air), and / or from radioactive decay of U and Th. Assuming that trapped air is not fractionated in $^{20}$Ne/$^4$He results in a contribution of <15% of atmospheric $^4$He for grain #126-05 (and even less for the other grains). Therefore, the majority (>85%) of the non-cosmogenic $^4$He is radiogenic. Diffusion of cosmogenic and radiogenic

He in olivine at typical equilibrium temperatures in the asteroid belt (~170 K) is very low, so that no diffusion correction is necessary, even at Ga-scales (Trull et al., 1991). We can thus determine a simple U,Th-He retention age, i.e., the time it would take to accumulate the radiogenic $^4$He concentration from the radioactive decay of U and Th present in the sample. A retention age shorter than the age of the solar system is indicative of a loss of radiogenic $^4$He at some point, e.g., due to impact shock-heating, or protracted solar heating. If that He-loss was complete, the retention age dates the (end of the) He-loss event. If not, the U,Th-He age is an upper limit to the age of the (last) He-loss event. Since it is not possible to determine if the He-loss was complete, we consider *every* U,Th-He age given here to be an upper limit age. For this reason, we also do not correct for atmospheric $^4$He, because such a correction (<15%) would not lead to a qualitatively different result.

To calculate U,Th-He retention ages, we also need U and Th concentrations. These were not measured directly, to maximize the mass available for noble gas analyses, but as mentioned the size and chemistry of the grains suggest they are fragments of chondrules. Individual chondrules of the CV$_{ox}$ chondrite Allende have bulk U concentrations of 14-23 ppb and a Th/U ratio of ~3.5 (Amelin and Krot, 2007), which is also compatible with the respective values in bulk CV chondrites (Wasson and Kallemeyn, 1988). However, Kööp and Davis (2012) found that U is inhomogeneously distributed *within* chondrules (of an LL3 chondrite): most of the U is concentrated in the mesostasis (up to 150 ppb U), leaving only about 5±5 ppb U in olivine. Although we refer to the grain mineralogy as "olivine" here, we only did SEM-EDS spot analyses of the grains (see section 2.2.), and thus cannot exclude that mesostasis or other minerals were present as minor phases in the grains. We find it unlikely, however, that all grains would be systematically depleted in mesostasis compared to a chondrule. To alleviate the concern that we might be underestimating our U,Th-He ages, we will calculate them for both 14-23 ppb U (the "nominal age") and 5 ppb U (the "maximum age").

The nominal ages of five of the grains are roughly consistent with each other at <400-700 (Table 1, Fig. 4). The maximum ages for these five grains are approximately <1.5-1.8 Ga (see Fig. S4 in the Supplementary Material). Grain #126-02, with a higher nominal age of <2.5–3.5 Ga (and a maximum age in excess of the age of the Solar System), might have had a higher fraction of (U-enriched) mesostasis than the other grains, and / or was only partially degassed. The consistent, short U,Th-He ages in all but one grain seem to indicate that the parent body of Khatyrka experienced a strong gas-loss event, with a nominal age of <600 Ma (the youngest upper limit age in the set, rounded to a single digit) and a maximum age of <1.8 Ga.

## 4. DISCUSSION

### 4.1. An improved chronology for the Khatyrka meteorite

Prior to this work, only two points in the history of Khatyrka were known: the Cu,Al-alloys formed at least 2-3 Ma after the condensation of the first solids in the solar system (i.e., less than ca. 4563 Ma ago; Amelin and Krot, 2007), and fragments of the meteorite were eventually deposited in a river sediment in the Far East of Russia, >7 ka ago (MacPherson et al., 2013). Now, we have established at least two additional events: (1) The time when the Khatyrka meteoroid was ejected from its parent asteroid, most likely 2-4 Ma ago; (2) The age of a strong shock event, which led to the loss of radiogenic $^4$He, about <600 Ma (and perhaps up to <1.8 Ga) ago. Since the record of cosmogenic $^3$He is undisturbed (see section 3.1., Figs. 1 and 2), the He-loss event must have happened *before* the start of cosmic-ray exposure (this also excludes that the shock event recorded in Khatyrka is recent, e.g., from anthropogenic activity). Stoeffler et al. (1991) show that radiogenic gas loss in ordinary chondrites starts at shock stage S3 (10 GPa), and is complete only when shock stage S5 is reached (35 GPa). Sharp and de Carli (2006) suggest that the pressures associated with shock stages are likely overestimated by Stoeffler et al. (1991) by a factor of two. Conservatively, the shock event experienced by Khatyrka must thus have reached a pressure of at least 5–18 GPa, and probably close to the upper end of that range given that most Khatyrka grains analyzed have been thoroughly reset. This range is consistent with the pressure >5 GPa deduced for Khatyrka (Hollister et al., 2014). Lin et al. (2017) suggested an S4 shock stage for Khatyrka. According to Stoeffler et al. (1991), an S4 shock stage entails a post-shock temperature increase of only 250–350 °C, much lower than the minimum shock temperatures recorded in some minerals from Khatyrka (>1200°C). However, higher post-shock temperatures are achievable if the target material has a significant porosity (e.g., Schmitt, 2000). Oxidized CV chondrites have indeed a relatively high porosity of 21.8±1.7% (Consolmagno et al., 2008). The shock temperature expected for a carbonaceous chondrite with a grain density of 3.3 g/cm³, an initial porosity of 22% (compressed to zero above 5 GPa), a characteristic heat capacity of 1 J/gK, and a pre-shock temperature of -100 °C (~170 K), is ~-40 °C for 5 GPa and ~1500 °C for 18 GPa (Sharp and de Carli, 2006). An impact shock at the upper end of this range can therefore explain both the He-loss and shock temperatures >1200 °C. Therefore, the event that degassed the Khatyrka olivine grains might well have been the same as the shock event that formed the second generation of quasicrystals (Lin et al. 2017).

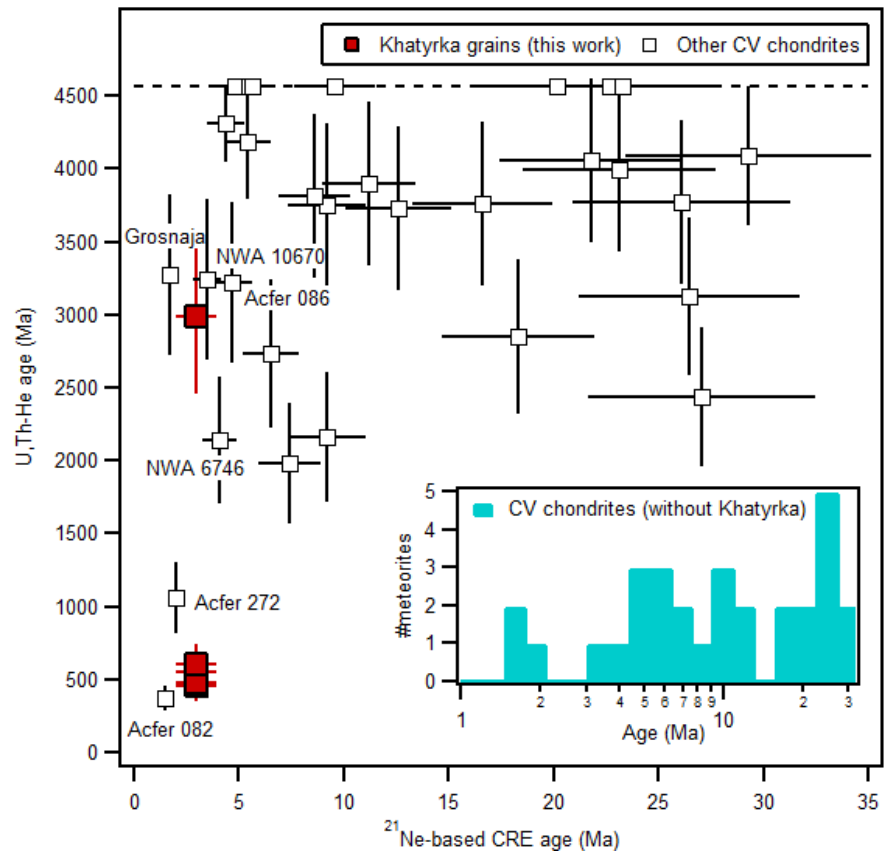

*Fig. 4: Cosmic histories of CV chondrites.* Main: U,Th-He retention age vs. ($^{21}$Ne-based) CRE age for all CV chondrites with published bulk He, Ne, Ar contents (all data points are simple averages of all samples measured for a meteorite; data compiled from Choi et al., 2017; Leya et al., 2013; Schultz and Franke, 2004), and Khatyrka. Meteorites mentioned in the main text are labeled. The U,Th-He ages (calculated here for the range U = 14-23 ppb and Th/U = 3.5) are capped at the age of the solar system (long-dashed line). All CRE ages are nominal values (given uncertainty = 20%). For the full data set, see Table S1 in Supplementary Material. Inset: CRE age histogram for all CV chondrites, with potential peaks at 2 Ma, 5 Ma, 10 Ma and 23 Ma.

## 4.2. Is Khatyrka related to other CV chondrites?

By comparing the CRE and U,Th-He ages of Khatyrka with their counterparts in other CV chondrites, we can potentially identify meteorites which experienced a similar cosmic history and might thus be "source-paired" with Khatyrka. This could lead to the discovery of more meteorites with Cu,Al-alloys and quasicrystals, or at least the precursor materials from which they may have formed. Out of 428 CV chondrites listed in the Meteoritical Bulletin Database (as of September 2017), only 30 have complete He, Ne, Ar records allowing the determination of both $^{21}$Ne-based CRE ages and U,Th-He retention ages (Fig. 4; Supplementary Table S1; data from Choi et al., 2017; Leya et al., 2013; Schultz and Franke, 2004). There are six CV chondrites with CRE ages that are roughly compatible with Khatyrka: Acfer 082, Acfer 086, Acfer 272, Grosnaja, NWA 6746, and NWA 10670. They all have higher U,Th-He retention ages than Khatyrka (even if we consider only the maximum age of 1.8 Ga), except for Acfer 082 and Acfer 272. These two meteorites, however, are thought to have lost most of their (cosmogenic and radiogenic) He through intense solar heating during CRE (Scherer and Schultz, 2000). In contrast, Khatyrka has retained its cosmogenic He, as discussed in sections 3.1 and 3.2. Therefore, the shock age of five of the six Khatyrka olivines seems to be unique (at least so far) among CV chondrites, even if we use the maximum age of 1.8 Ga. This only adds to the "uniqueness" of Khatyrka already suggested by the exotic Cu,Al-alloys, quasicrystals, and the unusual achondritic lithology found in the Florence sample. In summary, so far, it seems that Khatyrka does not have a potential "peer" among the other CV chondrites.

## 4.3. Candidate parent asteroids for Khatyrka

Before being delivered to an Earth-crossing orbit, meteoroids spend most of their CRE time (typically a few 10 Ma for stony objects) as small bodies in the asteroid belt. Their orbits evolve slowly due to gravitational encounters and the effect of non-gravitational forces, most notably, the Yarkovsky effect (Vokrouhlicky et al., 2015). Eventually, they reach an orbital resonance which ejects them from the asteroid belt, e.g., the 3:1 resonance at 2.50 au, where the meteoroid orbits the Sun exactly three times for every orbit of Jupiter. In Fig. 5 (top panel), we show the position (inclination vs. semi-major axis) of the seven most important resonances in the asteroid belt (Gladman, 1997), together with the positions of the largest asteroids (>30 km), for which non-gravitational forces are negligible, between 2.2 and 3.5 au (all asteroid data are from http://asteroid.lowell.edu). K-type asteroids, highlighted in green in Fig. 5, have long been associated with the CV, CK and CO type chondrites, based on similar reflectance spectra in the visible and near-infrared (Bell, 1988; Burbine et al., 2001; Cloutis et al., 2012).

The distance the meteoroid migrated in the asteroid belt (prior to ejection by a resonance) can be estimated from the CRE age, the meteoroid size, and some reasonable assumptions on meteoroid density and thermal properties, which, in combination, determine the strength of the non-gravitational forces altering the orbit. This migration distance can then be used to identify possible parent asteroids for Khatyrka. It should still be considered an approximation, because the meteoroid might also spend part of its CRE on an Earth-crossing orbit (after ejection from the resonance), and because its initial ejection from the parent asteroid will also result in an orbit slightly different from the one of its parent (e.g., an orbital velocity change of ~50 m/s, about the escape velocity of a ~100 km asteroid, results in a semi-major axis change of up to ~0.01-0.02 au). Using a simple model (Meier et al., 2017), we track the orbital migration of meteoroids with radii of 0.5, 1, 1.5, 3, and 5 meters, with corresponding CRE ages of 2.5, 2.5, 3.5, 5, and 6 Ma, respectively, in the vicinity of the 3:1, 5:2, and $v_6$ resonances (asteroids in the secular $v_6$ resonance have an orbital precession rate close to the one of Saturn). These are the only three resonances which can plausibly deliver the Khatyrka meteoroid to an Earth-crossing orbit within the short CRE age range of 2-4 Ma (Gladman, 1997). While the large K-type Eos family is overlapping the 9:4 resonance, the median time for meteoroid delivery to Earth through that resonance is about 90 Ma (Glad-man, 1997). As a consequence, the delivery of meteoroids from the 9:4 resonance (and the K-type Eos family located there) is very inefficient (di Martino et al., 1997). For a surface thermal conductivity $K = 1.5$ Wm$^{-1}$K$^{-1}$ (measured directly in CV chondrites; Opeil et al., 2012), the model yields migration distances on the order of 0.02-0.03 au in the vicinity of the 3:1, 5:2, and $\nu_6$ resonances. This distance is nearly independent of meteoroid size: despite the higher migration rate of small meteoroids, the higher production rate of cosmogenic nuclides in small meteoroids requires that their measured inventory of cosmogenic noble gases must have been accumulated in less time, thereby limiting the distance they can migrate.

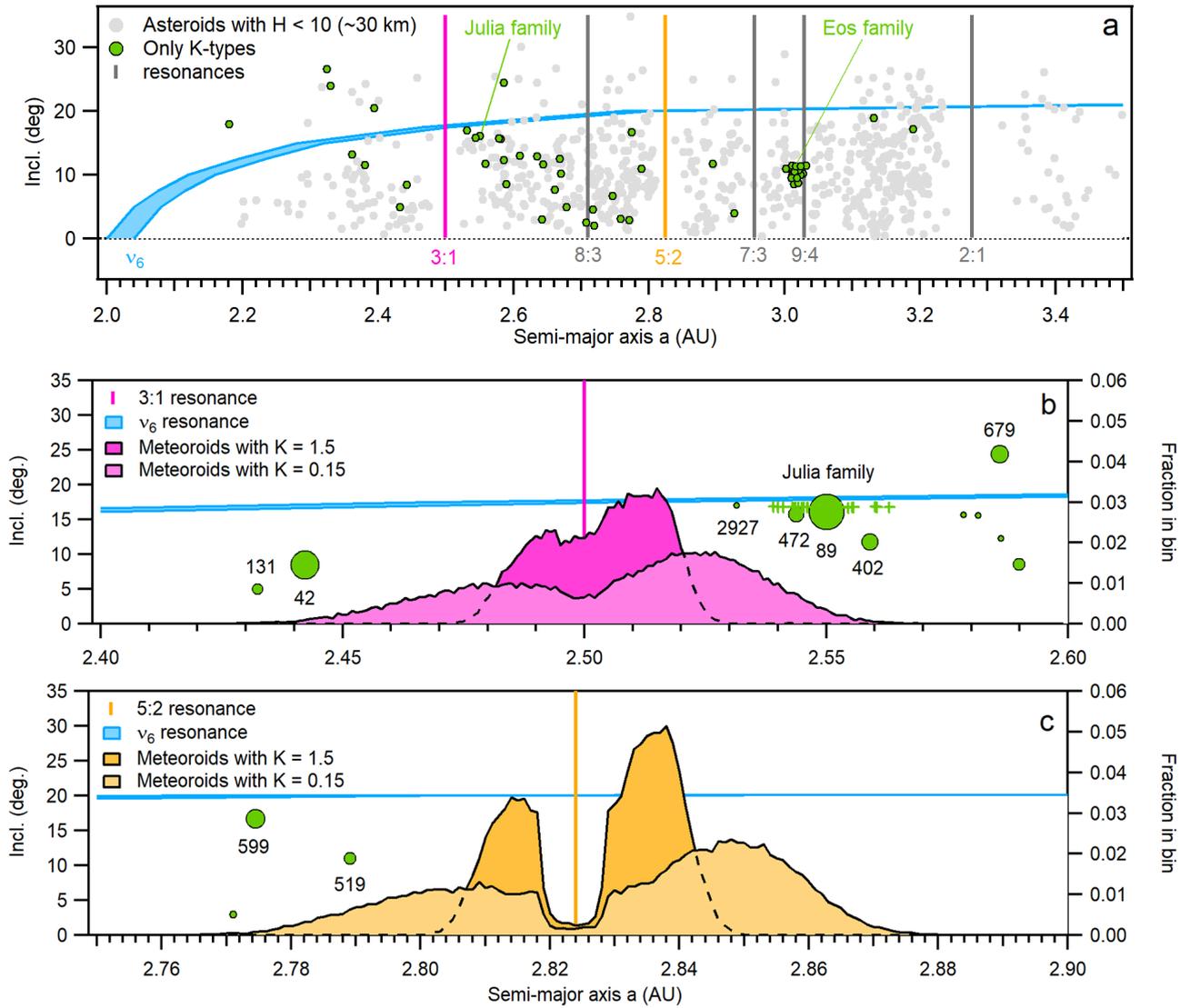

*Fig. 5: Potential parent asteroids for Khatyrka. Distribution of large asteroids (gray) and K-type asteroids (green) in the asteroid belt in inclination vs. semi-major axis space (top panel, a). Also shown are the positions of the seven most important resonances, and the position of two asteroid families discussed in the main text. The lower two panels show the cumulative probability density distribution (bin size 0.001 au, bin fraction given on right axis) of the initial semi-major axis of $5×10^4$ modeled Khatyrka meteoroids ($10^4$ for each of the five radii tested), for two surface thermal conductivities, 1.5 $Wm^{-1}K^{-1}$ (dark shades) and 0.15 $Wm^{-1}K^{-1}$ (light shades), in the vicinity of two orbital resonances which could have delivered the Khatyrka meteoroid within 2-4 Ma to Earth, 3:1 (middle panel, b) and 5:2 (bottom panel, c). Large K-type asteroids in this region are shown as green circles (with the symbol size proportional to the diameter, and the asteroid number given for the largest of them). Smaller members of the Julia family are shown as green crosses.*

In Fig. 5 (middle and lower panels), we show probability density distributions for the initial semi-major axes of the modeled meteoroids (dark shades: $K = 1.5$ Wm$^{-1}$K$^{-1}$). As there are no large K-type asteroids within 0.02-0.03 au of the three resonances, the Khatyrka meteoroid likely had a higher migration rate (although, as mentioned above, the initial "kick" at ejection from the parent asteroid might have added at least another ~0.02 au). Higher migration rates are possible if the surface is partially dust-covered and/or strongly fractured, which reduces the surface thermal conductivity, resulting in a stronger Yarkovsky effect (Vokrouhlicky et al., 2015). Surface thermal conductivities can vary by about three orders of magnitude between the very fine-grained lunar regolith ($K$ ~$10^{-3}$ Wm$^{-1}$K$^{-1}$) and bare rock ($K$ ~1 Wm$^{-1}$K$^{-1}$; Rubincam, 1995). As shown in Fig. 5, a reduction of the surface thermal conductivity by an order of magnitude (to $K = 0.15$ Wm$^{-1}$K$^{-1}$; light shades) allows the Khatyrka meteoroid to migrate a distance of 0.05-0.06 au in 2-4 Ma, and thus to reach the 5:2 resonance from asteroids 519 Sylvania and 599 Luisa (the latter asteroid has a spectrum closely matching the one of the CV chondrite Mokoia; Burbine et al., 2001). No large K-type asteroids are found within 0.05-0.06 au of the $\nu_6$ resonance (thus it is not shown as a separate panel in Fig. 5), but there are several large K-type asteroids within that distance of the 3:1 resonance. Perhaps most interestingly, the ~145 km diameter asteroid 89 Julia (the second-largest K-type asteroid after 15 Eunomia), has a compact (i.e., young) cratering family associated to it (Nesvorny et al., 2015). Asteroid 89 Julia and its family are also located relatively close to the $\nu_6$ resonance (both in semi-major axis and inclination), providing an additional channel through which meteoroids could be delivered to Earth-crossing orbits. Nesvorny et al. (2015) do not give an age for the Julia family, but a rough upper limit (ignoring the initial velocity distribution imparted on the fragments in the collision) can be determined from the largest distance between a small (km-sized) family member and 89 Julia, about 0.01 au. A characteristic migration rate for a km-sized K-type asteroid, at 2.5 au, is within a factor of a few of $10^{-4}$ au/Ma, which thus implies an age of a few 100 Ma, compatible with the <600 Ma age of the event that reset the U,Th-He clocks in Khatyrka.

Interestingly, asteroid 89 Julia displays an additional peculiarity, strengthening its possible connection to Khatyrka: while it fits the typical K-type spectrum well at short wave-lengths, it shows a steep red slope long-ward of 1 μm, and a plateau above 1.5 μm (Birlan et al., 2004). This was recognized already by Gaffey et al. (1993), who suggested that a contribution from a Ca-rich clinopyroxene lithology could explain 89 Julia's peculiar spectrum (see also Cloutis, 2002; while Gaffey et al. 1993 classified 89 Julia as an "ungrouped S", its shallow absorption at 1 μm is a better fit to the definition of the K-type asteroids. Bus and Binzel (2002) subsequently classified it as K-type). The Florence Sample of Khatyrka, in which the first quasicrystal was identified (Bindi et al., 2009), as well as some of the grains recovered in the 2011 expedition, represent an achondritic intergrowth of Cu,Al-alloys, olivine, spinel and diopside-hedenbergite – and the latter is indeed a Ca-rich clinopyroxene (MacPherson et al., 2013). If the unusual Florence Sample represents an important lithology on the Khatyrka parent asteroid, we would expect this asteroids' reflectance spectrum to resemble the peculiar one of 89 Julia. While a detailed spectral reflectance study of the different Khatyrka materials – including the Cu,Al-alloys – in the visible and near-infrared (~0.5 – 2.5 μm) is beyond the scope of this work, we encourage further investigation of the issue.

## 5. CONCLUSIONS

The analysis of light noble gases (He and Ne) in six forsteritic olivine grains from the quasicrystal-bearing $CV_{ox}$ meteorite Khatyrka shows that the precursor meteoroid must have traveled interplanetary space for 2-4 Ma if it was smaller than 1.5 m in radius, or, less likely, >4 Ma if it was larger and the sample heavily shielded. While this CRE age is similar to CRE ages in a few other CV chondrites, the U,Th-He retention ages of five of the six grains are much shorter than in any other CV chondrite, and imply that Khatyrka experienced a strong shock event, approximately <600 Ma ago (assuming chondritic U,Th concentrations in our grains). The shock pressures needed to reset the U,Th-He clock to the extent observed (at least ~5-18 GPa), and the corresponding shock temperatures (up to ~1500 °C) expected in an oxidized CV chondrite of typical porosity, are consistent with the minimum pressure and temperature required to explain some mineralogical observations in Khatyrka (Hollister et al., 2014; Lin et al., 2017). While several large K-type asteroids near the 3:1 and 5:2 resonances are potential parents for Khatyrka, the asteroid 89 Julia and its associated collisional family stand out by uniting four properties: (1) a K-type reflectance spectrum matching, at short wavelengths, the CV classification of Khatyrka; (2) a peculiar reflectance spectrum at longer wavelengths, suggestive of the presence of a Ca-rich clinopyroxene lithology, reminiscent of the unusual achondritic lithology represented by the "Florence sample" of Khatyrka; (3) proximity to two efficient and fast orbital resonances, 3:1 and $\nu_6$, capable of delivering a meteoroid within the CRE age of 2-4 Ma to Earth; and (4) a strong shock event within the last few 100 Ma (the cratering event which led to the formation of the Julia asteroid family), matching the U,Th-He age of approximately <600 Ma of Khatyrka olivine grains.

*Acknowledgments*: MM thanks J. Gilmour for the hospitality during his visit to SEAES. MM was supported by a Swiss National Science Foundation (SNSF) Ambizione grant. LB was supported by the "Progetto di Ateneo 2015" issued by the Università di Firenze, Italy. PRH was funded by a grant from the Tawani Foundation. H. Busemann's work was supported by NCCR PlanetS funded by SNSF. The authors also thank Jean-Luc Margot, reviewers Gregory F. Herzog and Julia A. Cartwright, as well as two anonymous reviewers of an earlier version of the manuscript, for their comments.

*Author contributions*: MM carried out the noble gas analysis and the orbital evolution modeling, wrote the manuscript, and together with RW, LB, PJS, and H. Busemann, conceived and developed the project. PRH measured chemical compositions of the grains, and together with AIN carried out the CT scan. NHS transferred all grains with the micro-manipulator, MEIR determined NeH interferences, CM maintained and set-up the unique noble gas mass spectrometer ready for analysis, and together with H. Baur helped with the development of the new fitting routine.

## Tables

*Table 1: Volumes, masses, noble gases and U,Th-He ages of six Khatyrka olivine grains*

|  | #126-01 | #126-02 | #126-03 | #126-04 | #126-05 | #126-07 |
|---|---|---|---|---|---|---|
| Voxels* | 219±14 | 190±13 | 228±15 | 292±18 | 274±23 | 181±15 |
| Volume, $\mu m^3$ | 35800±2300 | 31200±2100 | 37400±2400 | 47900±3000 | 44900±3800 | 29700±2500 |
| Diameter, $\mu m$** | 40.9±0.9 | 39.0±0.9 | 41.5±0.9 | 45.1±0.9 | 44.1±1.2 | 38.4±1.1 |
| #Fo*** | 0.96 | 0.96 | 0.98 | 0.91 | 0.97 | 0.96 |
| Calculated mass, $\mu g$ | 0.119±0.008 | 0.104±0.007 | 0.125±0.008 | 0.160±0.010 | 0.150±0.013 | 0.099±0.009 |
| Above DL**** | $^3$He, $^{21}$Ne | $^{3,4}$He, $^{20,21,22}$Ne | $^3$He, $^{20,21,22}$Ne | $^{3,4}$He, $^{21,22}$Ne | $^{3,4}$He, $^{20,21,22}$Ne | $^3$He, $^{21}$Ne |
| $^3$He/$^4$He | (>0.025) | 0.0035±0.0006 | (>0.026) | 0.025±0.009 | 0.026±0.020 | (>0.018) |
| $^3$He$_{meas}$ = $^3$He$_{cos}$ | 5.41±0.51 | 5.33±0.46 | 5.30±0.41 | 6.13±0.46 | 5.60±0.55 | 4.50±0.49 |
| $^4$He$_{rad}$ = $^4$He$_{non-cos}$ | (<190) | 1510±250 | (<180) | 220±90 | 190±170 | (<240) |
| $^{20}$Ne/$^{22}$Ne | -- | 11.5±1.7 | 17.7±3.9 | (<6.58) | 10.7±1.6 | -- |
| $^{21}$Ne/$^{22}$Ne | (>0.87) | 0.086±0.016 | 0.20±0.06 | 0.64±0.18 | 0.14±0.03 | (>0.26) |
| $^{21}$Ne$_{meas}$ | 1.83±0.21 | 1.36±0.22 | 0.91±0.19 | 1.80±0.18 | 1.33±0.22 | 0.65±0.17 |
| $^{21}$Ne$_{cos}$ | 1.83±0.22 | 0.92±0.22 | 0.80±0.20 | 1.78±0.19 | 1.08±0.22 | (>0.60) |
| $^3$He/$^{21}$Ne$_{cos}$ | 2.95±0.35 | 5.80±1.38 | 6.67±1.65 | 3.45±0.33 | 5.18±0.96 | 7.20±2.10 |
| $^{20}$Ne$_{cos}$/$^4$He | (>0.0079) | 0.0006±0.0002 | (>0.0036) | 0.0067±0.0025 | 0.0046±0.0036 | (>0.0023) |
| R4, Ma (U=14 ppb) | (<590) | 3520 | (<560) | 680 | 590 | (<740) |
| R4, Ma (U=23 ppb) | (<370) | 2460 | (<350) | 420 | 370 | (<460) |

*Note that the number of voxels are rounded, i.e., are not integers, due to the surface-effect corrections. **The diameter is for a spherical grain with the same volume. ***#Fo is the (nominal) forsterite number. #Fo = 0.95±0.05 was used to calculate the mass. ****"Above DL" lists the isotopes which have been detected above their respective detection limits as discussed in the methods section and the Supplementary Text. Concentrations are given in units of $10^{-8}$ cm$^3$STP/g, uncertainties are 1σ. Values in brackets are upper or lower limits resulting from only one of the two involved isotopes being above DL. "--" is given for ratios where neither of the two involved isotopes was above DL. "R4" are the nominal radiogenic ages determined from $^4$He$_{rad}$, assuming a bulk U concentration of either 14 or 23 ppb (and Th/U = 3.5).

## SUPPLEMENTARY TEXT

*An improved noble gas data fitting routine for very low gas amounts*

In the unique (inhouse-built) "compressor-source" noble gas mass spectrometer at ETH Zurich (Baur, 1999), used for the analysis of very small amounts of He and Ne (>~10'000 atoms), the "memory" signal from He and Ne constantly increases over the course of an analysis. While this phenomenon exists in many static vacuum noble gas mass spectrometers, it is exacerbated by the use of the compressor source, which concentrates gas from the entire volume of the mass spectrometer into the volume of the ion source. For samples with gas amounts close to the detection limit, the change in signal from this "memory" over the duration of an analysis is much larger than the signal from the sample. Heck et al. (2007) developed a protocol in which the increase of the memory signals in the static vacuum of the mass spectrometer and the extraction line is first monitored over four or five cycles (after separation from the turbo-molecular pumps). A cycle is a sequence in which the ion beam currents of all species of interest, typically $HD^+$, $^3He^+$, $^4He^+$, $H_2O^+$, $^{20}Ne^+$, $^{21}Ne^+$, $^{22}Ne^+$, $^{40}Ar^+$, and $CO_2^+$, are measured in peak jumping mode on a single detector (either a channel electron multiplier or a Faraday cup). Then, without opening or closing any valves (which might release additional gas into the spectrometer), the laser is fired at the sample, melting or in some cases completely vaporizing it, and releasing all He and Ne in the process. After extraction (referred to as "time zero"), another five to ten cycles are measured. In the cycles before the extraction, the gas in the spectrometer consists exclusively of memory gas, but after the extraction, the sample gas has been added. In the protocol developed by Heck et al. (2007), the signals measured in the four or five pre-extraction cycles are extrapolated forward in time to "time zero", while the signals measured in the post-extraction cycles are extrapolated backward in time to the same moment. The signal contribution from the sample gas is then defined as the difference between the values of the two extrapolations at time zero. The uncertainty of the sample signal is the quadratically added error of the two extrapolations at time zero. A wide variety of functions can be chosen to do these extrapolations, e.g., linear, quadratic, exponential or double exponential (a superposition of two exponential) functions. Whenever possible, the same type of function is used for the same ion species throughout an analysis session, unless the extrapolation obviously fails to provide a physically meaningful fit to the data. To determine the "detection limit", i.e., the smallest gas amount that can be reliably resolved from the background noise in this type of analysis, a series of "blank runs" is measured. These are identical to the normal sample runs, except that the laser is either not operated at all (a so-called "cold blank") or aimed at an empty spot on the Al sample holder (a so-called "hot blank"). The scatter of the gas amounts derived from the blank measurements, which typically scatter around zero regardless of their type, is then used to determine the detection limit, defined as two standard deviations of the blank scatter above the average blank value. A "detection" of gas for a specific species is only claimed for samples where the 1σ lower limit of the measured value is above the detection limit. This approach has allowed to reliably measure very small gas amounts in a variety of samples.

Nevertheless, this approach has its shortcomings. This can be exemplified using blank measurements, in which generally no gas is released at time zero. A single fit running through all cycles (both from before and after time zero) provides a better (i.e., more precise) estimate of the absolute signal intensity at time zero than the "classical" approach used above. First, this is because the available data set is larger than in the case of the two individual fits. Second, a fitting function is usually constrained best near the center of the data distribution. However, we are not interested in the centers of the pre-run and the post-run sections, and only really care about time zero, a point which is at the edge of each of the two time domains, i.e., where the fit functions are likely to provide a poor estimate. It would be better if the fitting function had "time zero" close to the center of its data distribution. This is possible when using a fitting function with a "step" added at time zero. By using a single function for the entire analysis run, we assume that the rate of signal change does not change at time zero, for which there is no physical justification if significant gas amounts are released by laser heating. We discuss this issue further below. In addition, and in spite of the concept of negative gas amounts released at time zero not being a physical one, the fitting algorithm is allowed to assume negative values for the step height. This is to allow for the statistical nature of the approach, e.g., due to counting statistics on individual data points, and not to bias the blanks to positive values. For a linear fit, the step function has the following form:

$s(t) = P_0 + P_1 \times 0.5 \times (1 + sign(t)) + P_2 \times t$

Here, s is the signal at time-index t, while $P_0$, $P_1$, and $P_2$ are fitting parameters (y-axis intercept, step-size, and slope, respectively), sign(t) is the "sign" function, which is -1 if t (time) is <0 (prior to time zero), and +1 if t is >0 (after time zero). Therefore, the value of the $P_1$-term ($P_1 \times 0.5 \times (1 + sign(t))$) is zero before time zero, and $P_1$ after that. Hence, $P_1$ corresponds to the signal increase due to the sample gas released at time zero (when the laser was fired). For non-linear functions, the P1-term is added analogously. Compared to the "classical" approach, this reduces the number of fitting parameters in the linear fit from four to three, from six to four in the quadratic and exponential fits, and from eight to five in the double-exponential fit. Furthermore, time zero is now close to the center of the distribution of values, which can be optimized by measuring the same number of cycles before and after time zero. The uncertainty on $P_1$ (determined by standard statistical methods and not to be confused with the error of the fit function at time zero) is the uncertainty of the signal change due to gas release at time zero. Since this value will still scatter around zero for cold and hot blanks, we still have to determine a detection limit using a suite of hot and cold blanks, but since their scatter (and their individual uncertainties) is reduced in the new fitting approach, this detection limit will, typically, be lower than in the "classical" approach.

The new fitting routine also has its limitations. For most species of interest, the rate of signal change at any time depends on the total signal of that species. Consequently, if the signal jumps due to the sample gas added at time zero is too large, the slope of the memory increase before and after time zero will be different, in which case the step function fit will be systematically wrong compared to the classical approach, which uses two independently fitted slopes. Therefore, the step function fit can only be used for very low gas amounts, which do not change the slope of the memory increase significantly. In particular, the step function cannot be used to fit a calibration run (an analysis of a known quantity of gas used to determine the isotope-specific sensitivity of the mass spectrometer) because the slope after time zero is negative for the large gas amounts introduced in a calibration run. Finally, and as a higher order effect, if laser heating releases a large amount of gas species other than a particular one of interest (possibly one that is not monitored), the slope of the memory signal increase of the species of interest (which also depends on the total pressure in the mass spectrometer) might also change at time zero. Therefore, the step function should only be used for relatively "clean" samples which release very small gas amounts. As a general rule, the signal jump at time zero should be smaller than the total memory increase of a species of interest over the course of an analysis. In summary, the step function cannot replace the classical approach, but it can complement and improve it for very low gas amounts.

# SUPPLEMENTARY TABLES

*Supplementary Table S1*

| Meteorite (CV) | $^3$He | $^4$He | $^{20}$Ne | $^{21}$Ne | $^{22}$Ne | $^{36}$Ar | $^{38}$Ar | $^{40}$Ar | Ref. | $^4$He$_{nc}$ | $^{21}$Ne$_c$ | R4, 14 | R4, 23 | T21 |
|---|---|---|---|---|---|---|---|---|---|---|---|---|---|---|
| Acfer 082 | 1.41 | 152 | 2.97 | 0.49 | 0.77 | 13.9 | 2.71 | 1070 | 1 | 144 | 0.48 | 453 | 279 | 1.5 |
| Acfer 086 | 10.7 | 1690 | 2.75 | 1.53 | 1.98 | 11.0 | 2.34 | 2910 | 1 | 1680 | 1.52 | 3770 | 2670 | 4.7 |
| Acfer 272 | 0.25 | 440 | 3.12 | 0.66 | 1.03 | 29.6 | 5.76 | 383 | 1 | 439 | 0.65 | 1300 | 820 | 2.0 |
| ALH84028 | 41.6 | 2700 | 9.98 | 7.11 | 8.52 | 16.6 | 4.14 | 2360 | 1 | 2450 | 7.09 | 4620 | 3500 | 21.8 |
| ALH85006 | 5.95 | 3160 | 9.00 | 1.45 | 2.42 | 24.0 | 4.68 | 1050 | 1 | 3120 | 1.42 | *4570 | 4050 | 4.4 |
| ALHA81003 | 26.7 | 2269 | 5.05 | 4.13 | 4.88 | 8.07 | 1.96 | 2700 | 1 | 2110 | 4.12 | 4290 | 3160 | 12.6 |
| Allende | 7.63 | 2830 | 5.48 | 1.76 | 2.28 | 16.9 | 3.30 | 2310 | 1 | 2780 | 1.75 | *4570 | 3790 | 5.4 |
| Arch | 112 | 273000 | 785 | 8.50 | 71.5 | 95.3 | 19.1 | 350 | 1 | 273000 | 6.57 | *4570 | *4570 | 20.2 |
| Axtell | 22.4 | 2290 | 8.56 | 5.41 | 6.23 | 10.2 | 2.59 | 211 | 1 | 2150 | 5.41 | 4330 | 3200 | 16.6 |
| Bali | 26.2 | 2740 | 18.7 | 9.58 | 11.3 | 30.7 | 6.67 | 1530 | 1 | 2580 | 9.55 | *4570 | 3610 | 29.3 |
| Denman 002 | 20.0 | 2260 | 5.71 | 3.01 | 3.77 | 6.95 | 1.81 | 752 | 1 | 2140 | 3.00 | 4320 | 3190 | 9.2 |
| Efremovka | 9.93 | 1050 | 12.3 | 3.03 | 4.10 | 246 | 45.8 | 1390 | 1 | 990 | 3.01 | 2600 | 1720 | 9.2 |
| Grosnaja | 1.15 | 1730 | 6.93 | 0.56 | 1.30 | 25.3 | 4.95 | 1040 | 1 | 1720 | 0.54 | 3820 | 2720 | 1.7 |
| Kaba | 4.15 | 2310 | 11.9 | 3.70 | 5.25 | 39.4 | 8.39 | 597 | 1 | 2280 | 3.66 | 4460 | 3330 | 11.2 |
| Leoville | 14.1 | 2280 | 11.7 | 2.84 | 4.05 | 226 | 41.8 | 1790 | 1 | 2200 | 2.81 | 4380 | 3250 | 8.6 |
| Mokoia | 57.9 | 97600 | 334 | 3.91 | 29.6 | 23.5 | 4.80 | 1410 | 1 | 97500 | 3.12 | *4570 | *4570 | 9.6 |
| Mundrabilla 012 | 15.4 | 1350 | 2.22 | 2.14 | 2.56 | 1.67 | 0.72 | 941 | 1 | 1330 | 2.13 | 3240 | 2220 | 6.5 |
| Nova 002 | 1.86 | 899 | 9.53 | 2.42 | 3.45 | 183 | 35.1 | 617 | 1 | 888 | 2.40 | 2390 | 1570 | 7.4 |
| NWA 10670 | 6.01 | 1730 | 3.28 | 1.15 | 1.61 | 2.86 | 0.65 | 167 | 2 | 1700 | 1.14 | 3800 | 2690 | 3.5 |
| NWA 3304 | 298 | 604000 | 1110 | 10.0 | 95.5 | 59.2 | 12.0 | 283 | 2 | 604000 | 7.41 | *4570 | *4570 | 22.7 |
| NWA 6743 | 59.2 | 1510 | 10.4 | 8.87 | 10.8 | 102 | 20.6 | 954 | 2 | 1150 | 8.84 | 2920 | 1960 | 27.1 |
| NWA 6746 | 5.59 | 1010 | 4.67 | 1.36 | 1.88 | 0.72 | 0.14 | 73.1 | 2 | 976 | 1.35 | 2570 | 1700 | 4.1 |
| QUE 93429 | 50.3 | 2460 | 10.5 | 8.53 | 9.93 | 17.8 | 4.79 | 1850 | 1 | 2160 | 8.52 | 4340 | 3210 | 26.1 |
| QUE 93744 | 52.1 | 1930 | 10.9 | 8.67 | 10.2 | 30.8 | 7.24 | 3400 | 1 | 1610 | 8.65 | 3670 | 2580 | 26.5 |
| RaS 221 | 25.1 | 1570 | 8.72 | 5.99 | 7.39 | 15.2 | 3.51 | 667 | 3 | 1410 | 5.97 | 3370 | 2320 | 18.3 |
| RaS 251 | 9.83 | 6420 | 6.29 | 1.59 | 2.37 | 4.94 | 1.34 | 2450 | 3 | 6360 | 1.57 | *4570 | *4570 | 4.8 |
| Tibooburra | 42.2 | 2630 | 11.3 | 7.54 | 9.11 | 88.5 | 17.4 | 171 | 1 | 2380 | 7.52 | 4560 | 3430 | 23.1 |
| Vigarano | 8.75 | 4940 | 17.1 | 1.85 | 3.32 | 43.3 | 8.39 | 2120 | 1 | 4880 | 1.81 | *4570 | *4570 | 5.5 |

| | | | | | | | | | | | | | |
|---|---|---|---|---|---|---|---|---|---|---|---|---|---|
| Y 86009 | 12.9 | 23500 | 147 | 2.16 | 13.6 | 58.0 | 11.2 | 799 | 1 | 23500 | 1.81 | *4570 | *4570 | 5.6 |
| Y 86751 | 27.7 | 21300 | 138 | 7.92 | 19.0 | 169 | 33.4 | 1210 | 1 | 21200 | 7.60 | *4570 | *4570 | 23.3 |

Compilation of noble gas data used to create Fig. 4. All concentrations are given in $10^{-8}$ cm$^3$STP/g. References (Ref.) are (1) the compilation by Schultz and Franke (2004), and research papers by (2) Choi et al. (2017) and (3) Leya et al. (2013). The cosmogenic $^{21}$Ne ($^{21}$Ne$_c$) was calculated by a two-component deconvolution as described in section 3.1. The non-cosmogenic $^4$He ($^4$He$_{nc}$) was calculated as follows: $^4$He$_{nc}$ = $^4$He$_{meas}$ − 6 × $^3$He$_{meas}$, unless $^3$He$_{meas}$ was larger than 7 × $^{21}$Ne$_c$, in which case that value was subtracted from $^4$He$_{meas}$ instead (based on Leya & Masarik, 2009). The U,Th-He ages (R4; in Ma) are given for U concentrations of 14 ppb and 23 ppb, in the third- and second-to last column, respectively (Th/U = 3.5; R4 ages are caped at 4570 Ma). The $^{21}$Ne-based CRE age (T21; in Ma) in the last column assumes a uniform $^{21}$Ne production rate of $0.326 \times 10^{-8}$ cm$^3$STP/gMa.

*Supplementary Table S2*

| Grain | O (%) | Na (%) | Mg (%) | Al (%) | Si (%) | Ca (%) | Fe (%) | Total (%) | Mg# | Description |
|---|---|---|---|---|---|---|---|---|---|---|
| #126-01 | 45.91 | n.d. | 33.41 | n.d. | 19.14 | n.d. | 1.54 | 100 | 0.96 | Forsterite |
| #126-02 | 42.37 | n.d. | 35.29 | n.d. | 20.83 | n.d. | 1.51 | 100 | 0.96 | Forsterite |
| #126-03 | 52.44 | n.d. | 30.81 | n.d. | 16.1 | n.d. | 0.66 | 100 | 0.98 | Forsterite |
| #126-04 | 45.32 | n.d. | 32.05 | n.d. | 19.6 | n.d. | 3.03 | 100 | 0.91 | Forsterite |
| #126-05 | 51.01 | n.d. | 30.58 | n.d. | 17.48 | n.d. | 0.93 | 100 | 0.97 | Forsterite |
| #126-06 | 46.46 | n.d. | 30.08 | n.d. | 18.13 | n.d. | 5.34 | 100 | 0.85 | Forsterite |
| #126-07 | 55.21 | n.d. | 29.36 | n.d. | 14.34 | n.d. | 1.09 | 100 | 0.96 | Forsterite |
| Fo 95 | 44.49 | - | 32.10 | - | 19.52 | - | 3.88 | 100 | 0.95 | (theoretical) |

Qualitative chemical composition of the seven original grains determined by SEM-EDS. Grain #126-06 was not selected for analysis. All abundances have been renormalized to 100%. „n.d."= not detected.

**SUPPLEMENTARY FIGURES**

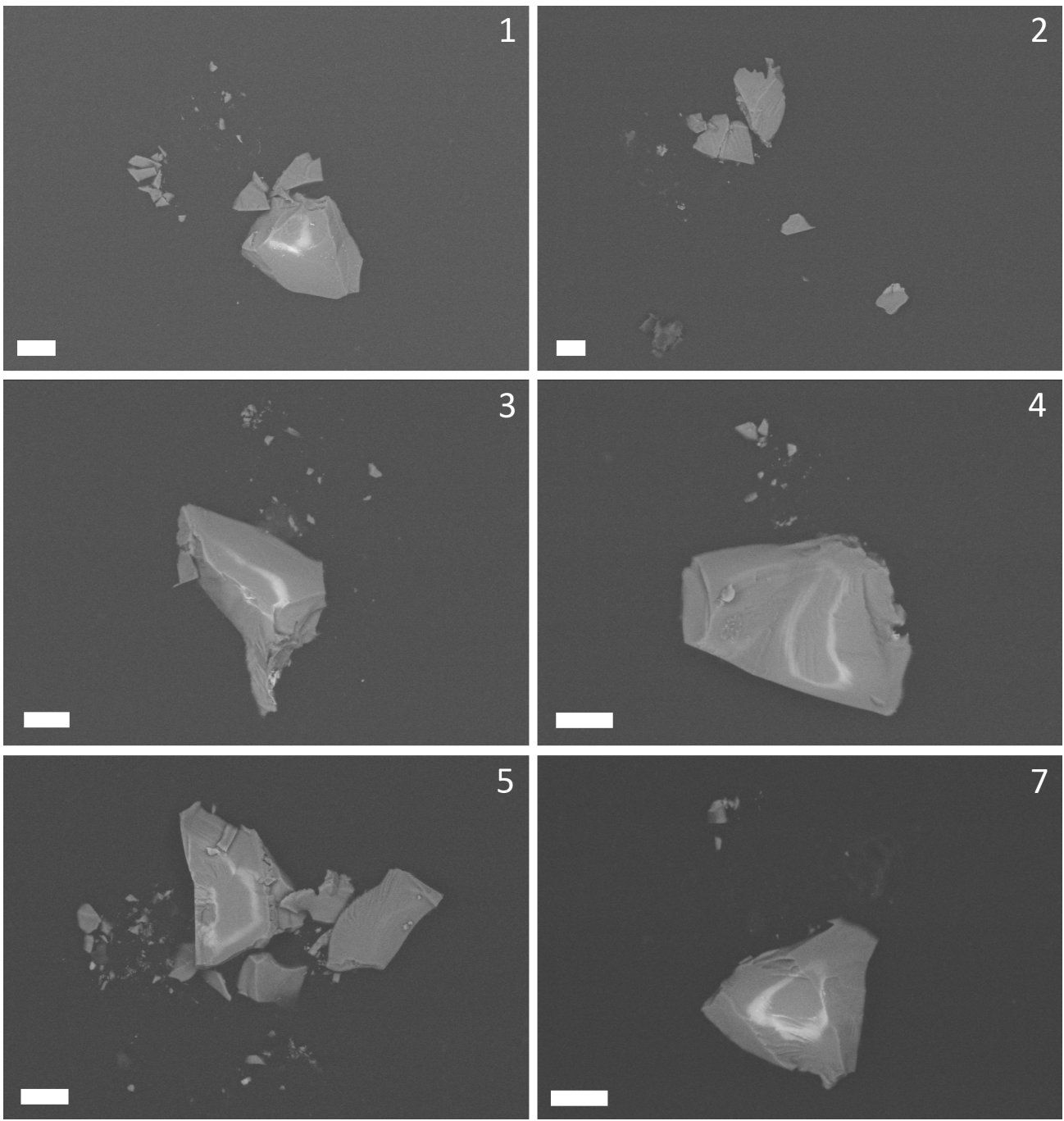

***Fig. S1: Grain BSE images.*** *All six Khatyrka olivine grains imaged by SEM (BSE) after nano-CT scanning. The white scale bar is 20 µm long in each image (Field Museum of Natural History, Chicago).*

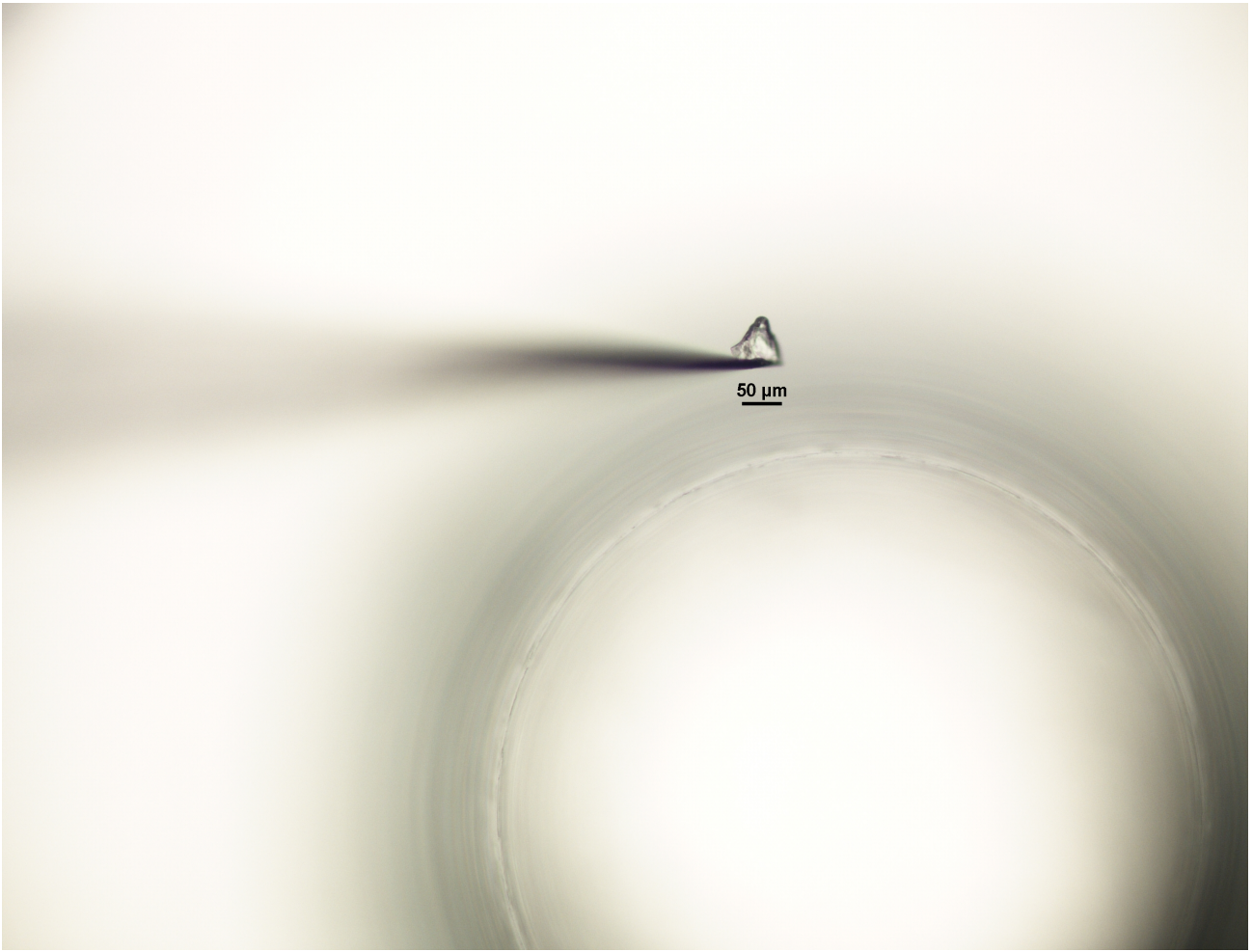

***Fig. S2: Khatyrka olivine grain on micro-manipulator needle.*** *Grain #126-07 on the tip of a micro-manipulator needle, during transfer from the carbon tape to the sample holder which was later used for noble gas analysis (SEAES, University of Manchester).*

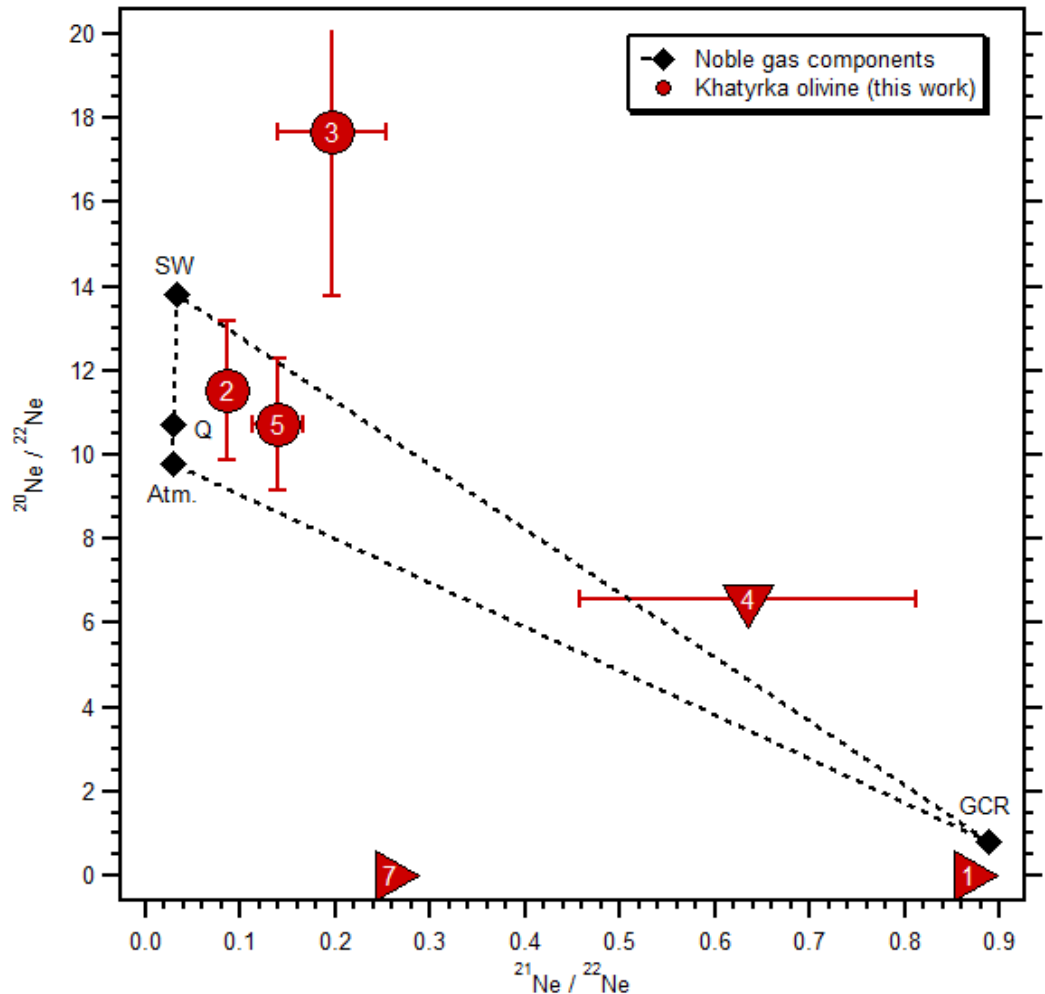

*Fig. S3: Three isotope diagram for Ne.* Neon isotope ratios ($^{20}Ne/^{22}Ne$ vs. $^{21}Ne/^{22}Ne$) for all six Khatyrka olivine grains. For grains #126-01 and -07, only $^{21}Ne$ was detected above the detection limit, and therefore, only a lower bound (2σ) on the $^{21}Ne/^{22}Ne$ ratio can be given – the data points (right-pointing triangles) are placed arbitrarily at $^{20}Ne/^{22}Ne = 0$. For grain #126-04, $^{20}Ne$ was not detected, therefore only an upper bound (2σ) on the $^{20}Ne/^{22}Ne$ ratio can be given (down-pointing triangle). Grain #126-03 plots nominally outside the field in which we would expect it (surrounded by dashed lines), although it is still less than 2σ away. Also shown are four noble gas components (black diamonds): the solar wind as measured by Genesis (SW; Heber et al., 2009), Q gases and the Earth's atmosphere (Ott, 2014), and cosmogenic Ne (GCR; Leya and Masarik, 2009).

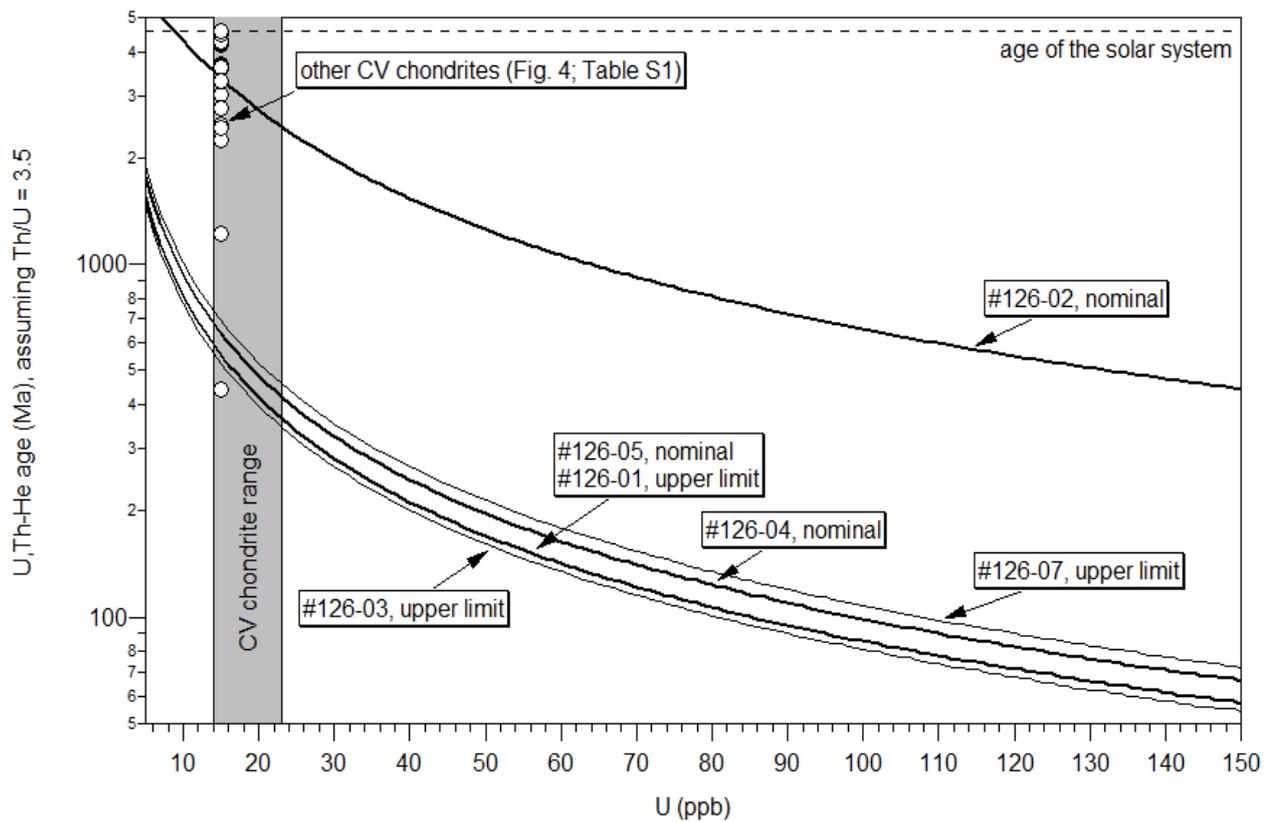

*Fig. S4: U,Th-He ages as a function of U concentration.* The U,Th-He age of all Khatyrka grains (black curves) is shown as a function of the U concentration in the range 5-150 ppb, assuming Th/U = 3.5. Also shown, for comparison, are the other CV chondrites (open circles; for 15 ppb U) as given in Fig. 4 and Supplementary Table S1.